\documentclass[aps,preprint,nofootinbib,showpacs]{revtex4}

\usepackage{amsmath,amssymb,bm}
\usepackage{yfonts}
\renewcommand\d{\partial}
\newcommand\+{\dagger}
\newcommand\tr{\mathop{\mathrm{tr}}}
\newcommand\LambdaQCD{\Lambda_{\rm QCD}}

\begin{document}

\preprint{INT-PUB 07-24}
\author{D.~T.~Son}
\affiliation{Institute for Nuclear Theory, University of Washington, 
Seattle, Washington 98195-1550, USA}

\author{M.~A.~Stephanov}
\affiliation{Department of Physics, University of Illinois, Chicago, 
Illinois 60607, USA}

\title{
%Nuclear matter in strong magnetic field
%and ferromagnetic quark matter
Axial anomaly and magnetism of nuclear and quark matter
}

\begin{abstract}

  We consider the response of the QCD ground state at finite baryon
  density to a strong magnetic field $B$. We point out the dominant
  role played by the coupling of neutral Goldstone bosons, such as
  $\pi^0$, to the magnetic field via the axial triangle anomaly.  We show
  that, in vacuum, above a value of $B\sim m_\pi^2/e$, a metastable
  object appears---the $\pi^0$ domain wall. Because of the axial anomaly,
  the wall carries a baryon number surface density proportional to 
  $B$. As a result,
  for $B\gtrsim 10^{19} ~\textrm G$ a stack of parallel $\pi^0$ domain
  walls is energetically more favorable than nuclear matter at the
  same density. Similarly, at higher densities,
  somewhat weaker magnetic fields of order 
  $B\gtrsim 10^{17}-10^{18}
  ~\textrm G$ 
  transform the color-superconducting
  ground state of QCD into new phases containing stacks of
  axial isoscalar ($\eta$ or $\eta'$) domain walls.  We also show that a
  quark-matter state known as ``Goldstone current state,'' in which a
  gradient of a Goldstone field is spontaneously generated, is
  ferromagnetic due to the axial anomaly.  We estimate the size
  of the fields created by such a state in a typical neutron
  star to be of order $10^{14}-10^{15} ~\textrm G$.

\end{abstract}

\maketitle

\section{Introduction}

There have been several studies of the structure of QCD vacuum in high
magnetic fields 
\cite{Shushpanov:1997sf,Kabat:2002er,Miransky:2002rp,Cohen:2007bt}.  
The typical strength of a magnetic field which would
change the structure of the QCD
vacuum is very high and can be estimated as
\begin{equation}\label{B10^20}
   B \sim \frac{m_\rho^2} e \sim 10^{20}~\textrm{G},
\end{equation}
where $m_\rho=770~\textrm{MeV}$ is the typical energy scale of QCD.
For example, the typical magnetic field that changes substantially
the chiral condensate is $(4\pi f_\pi)^2/e$ \cite{Shushpanov:1997sf}, 
which is of the same order as in Eq.~(\ref{B10^20}).
In Ref.~\cite{Kabat:2002er} it was argued that for
$B\gtrsim 10~\textrm{GeV}^2\approx 5\cdot10^{21}~\textrm{G}$ 
a condensate of spin-polarized $u\bar u$ pairs appear.
%or comparison the magnetic field in neutron
%tars is believed to not exceed $3\cdot10^{17}~\textrm{G}$.)

The behavior of nuclear matter in strong magnetic fields has been
studied more extensively.
The motivation for such studies is the high magnetic
field observed in magnetars~\cite{Duncan:1992hi}. On general grounds one
expects (see, e.g., Ref.~\cite{Broderick:2000pe}) that the
magnetic field affects significantly the structure of the matter
once the synchrotron (Landau level) energy $\sqrt{eB}$ is comparable
to the typical energy associated with charge excitations in the system,
such as, e.g., proton Fermi energies in nuclear matter.

The response of color-superconducting quark matter to a strong magnetic
field has also been studied
\cite{Alford:1999pb,Ferrer:2005vd,Ferrer:2006vw,Ferrer:2007iw,Ferrer:2006ie,Fukushima:2007fc,Noronha:2007wg}.
Similarly, in all mechanisms studied so far, the ground state is affected
above some value of the magnetic field determined by the
superconducting gap $\Delta$ and/or the chemical potential $\mu$.
For example, fields of order $\mu\Delta/e$ or higher are needed to destroy 
color superconductivity~\cite{Alford:1999pb}.

In this paper we show
that, due to the anomalous coupling of neutral pseudoscalar Golstone bosons 
to electromagnetism, the structure of the ground state is modified
at much lower values of the magnetic field.  In fact, these values
are parametrically lower than~(\ref{B10^20}) in the 
limit where the Golsdtone bosons become massless (e.g., the chiral limit).

For the low-density nuclear matter we find two scales of magnetic
field that are relevant (see Sec.~\ref{sec:pi0-domain-wall}):
\begin{equation}\label{B0}
  B_0 = \frac{3m_\pi^2}e, \qquad B_1 = 16\pi \frac{f_\pi^2 m_\pi}{em_N}\,.
\end{equation}
In particular, above $B_1$ nuclear matter is replaced by 
a different state.
The most striking feature of Eq.~(\ref{B0}) is that both
$B_0$ and $B_1$ \emph{vanish} in the chiral limit: when $m_\pi=0$,
the structure of nuclear matter
is altered at an arbitrarily small magnetic field! This is 
in sharp contrast to the previous estimates of
the critical magnetic field, Eq.~(\ref{B10^20}).

The state of QCD associated with scales (\ref{B0}) is a
$\pi^0$ domain wall---a configuration in which the local expectation
value of the $\pi^0$ field varies along the direction of the magnetic
field $\bm B$ over a scale of pion Compton wavelength. We show that
for $|\bm B|>B_0$ the domain wall becomes locally stable
(metastable).  

The central observation of this paper is that such a domain wall
carries nonzero surface baryon charge density proportional to $|\bm
B|$.  As we show, this is a consequence of the quantum axial anomaly---the
triangle anomaly involving the baryon, electromagnetic and neutral
axial currents.\footnote{The physics of triangle anomaly at
  finite
density has also received some interest recently, see, e.g.,
\cite{Son:2004tq,Metlitski:2005pr,Newman:2005as,Harvey:2007rd}.}
  When $|\bm B|>B_1$ the parallel stack of
such domain walls is energetically more favorable at $\mu\approx m_N$
than low density nuclear matter, as it carries less energy per
baryon.  That means nuclear matter turns into a
stack of $\pi^0$ domain walls at such large magnetic fields. For
larger magnetic fields this ``wall state'' should persist down to
chemical potentials $\mu \gtrsim m_N\,B_1/|\bm B|$.

We note right away that although both $B_0$ and $B_1$ 
vanish in the chiral limit $m_\pi\to0$
(with $B_0\ll B_1$), for the physical pion mass, these magnetic fields
are of order $10^{19}~\textrm{G}$, smaller than the QCD
scale~(\ref{B10^20}), but still much larger than 
the fields typical of magnetars.

The crucial role in our analysis is played by the Wess-Zumino-Witten
(WZW) term describing the anomalous interaction of the neutral pion
field with the external electromagnetic field, and a related pion
contribution to the baryon current.  For example, the WZW term
describes the anomalous $\pi^0\to2\gamma$ decay.  We review the
prerequisite basics of the WZW action in Sec.~\ref{sec:WZW}.  We then derive
the scales~(\ref{B0}) in Sec.~\ref{sec:pi0-domain-wall}.

In Sec.~\ref{sec:colorsup} we show that the same mechanism that leads
to the formation of $\pi^0$ domain walls in vacuum also 
operates in color-superconducting phases of QCD at high baryon densities.
Such phases could exist in the cores of dense neutron or quark stars.
The Nambu-Goldstone bosons associated with broken symmetries in these phases
are much lighter~\cite{inverse-ordering,SSZ} 
than $\pi^0$ in vacuum. As a result,
 in these phases, the domain walls appear spontaneously at
lower magnetic fields of order $10^{17}-10^{18}~\textrm{G}$, which decrease
with increasing $\mu$ due to the decrease of the Nambu-Goldstone boson masses.

Finally, in Sec.~\ref{sec:ferro} we consider another consequence of
the anomaly: the spontaneous generation of magnetization, i.e.,
ferromagnetism, in dense QCD matter. Ferromagnetism of nuclear and
quark matter, under various mechanisms,
has been discussed in the literature
\cite{Tatsumi:2000dv,Isayev:2003fz,Inui:2007zc,Ferrer:2007uw}.
It has been suggested that ferromagnetism may help explaining certain
features of magnetars~\cite{Bhattacharya:2007ud}.
We point out that for such magnetization to appear, it is sufficient
for a {\em pseudoscalar} Goldstone boson field to develop a nonzero average
spatial gradient.  Such a situation may indeed appear in the so-called
``Goldstone boson current'' phases of quark matter with mismatched
quark Fermi surfaces.  In the case when all gapless fermions are
electrically neutral, we show that the magnitude of the magnetization
is determined by the triangle anomalies.  We estimate this magnitude
in one particular scenario of Goldstone boson current in the
color-flavor-locked phase with neutral kaon condensation (CFLK$^0$
phase) to be of order $10^{16}~\textrm{G}$. Since only a finite (and
presumably small) region inside the neutron star is occupied by this
current phase, we estimate the typical magnetic field generated by
such a mechanism to be of order $10^{14}-10^{15}~\textrm G$.  If such a
mechanism indeed operates within the cores of some magnetars, it might
account for their unusually large magnetic fields.

\section{The WZW action in electromagnetic field}
\label{sec:WZW}

\subsection{SU(3) case}
\label{sec:su3}

We start from the SU(3) chiral perturbation theory, which describes the
octet of pseudoscalar Nambu-Goldstone bosons in terms of a 
$3\times3$ unitary matrix $\Sigma$
\begin{equation}\label{Sigma}
 \Sigma = \exp\left( \frac{i\lambda^a \varphi^a}{f_\pi}\right),
\end{equation}
where $\lambda^a$ are the 8 Gell-Mann matrices and
\begin{equation}
  \frac1{\sqrt2} \lambda^a\varphi^a =\left( \begin{array}{ccc}
    \frac{\pi^0}{\sqrt2}+\frac\eta{\sqrt6}& \pi^+ & K^+ \\
    \pi^- & -\frac{\pi^0}{\sqrt2}+\frac\eta{\sqrt6} & K^0 \\
    K^- & K^0 & -\frac{2\eta}{\sqrt6}
  \end{array} \right).
\end{equation}
Without the WZW term, the Lagrangian of the theory in an external 
electromagnetic field $A_\mu$ is
\begin{equation}\label{eq:sigma-lagrangian}
  {\cal L} = \frac{f_\pi^2}4 \tr D_\mu\Sigma^\+ D_\mu\Sigma + 
     \tr(M\Sigma+\textrm{h.c.}),
\end{equation}
where
\begin{equation}
  D_\mu\Sigma = \d_\mu\Sigma + ieA_\mu [Q,\, \Sigma],
\end{equation}
with $Q=\mathrm{diag}(2/3,-1/3,-1/3)$. The Lagrangian is invariant under 
global
SU(3)$_L\times$SU(3)$_R$ symmetry, and under the local U(1)$_Q$ subgroup
of this symmetry. Gauging the whole SU(3)$_L\times$SU(3)$_R$
in QCD is not possible due to the axial 
anomalies~\cite{anomaly}. 
%but the Lagrangian
%in (\ref{eq:sigma-lagrangian}) does not show that. 
The anomalies are captured by the 
%correctly reproduces anomalous current correlation functions is the
Wess-Zumino-Witten (WZW) term in the action~\cite{Wess:1971yu,Witten:1983tw}.
We introduce the standard notations,
\begin{equation}
  L_\mu = \Sigma\d_\mu\Sigma^\+, \qquad
  R_\mu = \d_\mu\Sigma^\+\Sigma .
\end{equation}
%The physics of the axial anomaly in the effective theory is
%represented by the WZW action. 
In the background of the external
electromagnetic field $A_\mu$ as well as an auxiliary gauge potential
$A_\mu^B$ coupled to baryon current, the WZW term is given
by~\cite{Wess:1971yu,Witten:1983tw,Kaymakcalan:1983qq,DGH}
\begin{multline}\label{SWZW}
  S_{\rm WZW}[\Sigma, A_\mu, A^B_\mu] = S_{\rm WZW}[0] 
  - \int\!d^4x\, A^B_\mu j_B^\mu
  + \frac{\epsilon^{\mu\nu\alpha\beta}}{16\pi^2} \int\!d^4x\,  
    \Bigl[ e A^\mu
    \tr(QL_\nu L_\alpha L_\beta + QR_\nu R_\alpha R_\beta)\\
  -ie^2 F_{\mu\nu}A_\alpha \tr(Q^2L_\beta +Q^2 R_\beta +
  \tfrac12 Q\Sigma Q\d_\beta\Sigma^\+
  -\tfrac12 Q\Sigma^\+ Q\d_\beta\Sigma ) \Bigr].
\end{multline}
Here $S_{\rm WZW}[0]$ is the WZW term without the gauge field (which
can be written in the form of a 
five-dimensional integral). The additional terms in (\ref{SWZW})
make the action invariant with respect to
local U(1)$_B$ and U(1)$_Q$ (baryon and electric charge) transformations.

The U(1)$_B$ transformation is not a part of the
SU(3)$_L\times$SU(3)$_R$ group and the fields $\Sigma$ do not
transform under it. However, the external U(1)$_B$ gauge potential
$A^B_\mu$ does couple to $\Sigma$ via the Goldstone-Wilczek baryon
current $j_B^\mu$~\cite{Goldstone:1981kk,Witten:1983tw}.  In the external
electromagnetic field, the conserved and gauge invariant baryon
current $j_B^\mu$ can be found using the ``trial and error'' gauging,
following Witten \cite{Witten:1983tw}
\begin{equation}
 % \label{eq:witten}
    j_B^\mu = -\frac1{24\pi^2}\epsilon^{\mu\nu\alpha\beta}\left\{
  \tr(L_\nu L_\alpha L_\beta)
%  + \frac{ie}{8\pi^2}
%\epsilon^{\mu\nu\alpha\beta}
- 3ie
\d_\nu[A_\alpha   
    \tr (QL_\beta +QR_\beta)]
\rule{0pt}{1em}\right\},
\label{jB-cons}
\end{equation}
or the ``covariant derivative'' gauging, following
Goldstone and Wilczek~\cite{Goldstone:1981kk}
\begin{equation}
  j_B^\mu = - \frac1{24\pi^2}\epsilon^{\mu\nu\alpha\beta}
\left\{
    \tr[(\Sigma D_\nu\Sigma^\+) (\Sigma D_\alpha\Sigma^\+) 
    (\Sigma D_\beta\Sigma^+)]
%  +\frac{ie}{16\pi^2}\epsilon^{\mu\nu\alpha\beta}
-\frac{3ie}{2}
F_{\nu\alpha}
  \tr[Q(\Sigma D_\beta\Sigma^+ + D_\beta\Sigma^\+\Sigma)]
\right\}.
  \label{jB-gi}
\end{equation}
In the form (\ref{jB-cons}) both terms are obviously conserved, but not
separately gauge invariant. In the form (\ref{jB-gi}) both terms
are obviously gauge invariant, but not separately conserved.
It can be checked that the two forms are equivalent.

\subsection{SU(2) case}
\label{sec:su2}

If one specializes to the SU(2) case [i.e., only $\varphi^1$, $\varphi^2$,
$\varphi^3$ are nonzero in Eq.~(\ref{Sigma})], then the 
previous formulas simplify.
We can write
\begin{equation}
  \Sigma = \frac1{f_\pi} (\sigma + i\tau^a\pi^a), 
  \qquad \sigma^2+\pi^a\pi^a=f_\pi^2,
\end{equation}
and $Q=t^3+1/6$ ($t^3=\tau^3/2$) to verify, e.g., 
that $\tr(Q\Sigma Q\d_\beta\Sigma^\+
-Q\Sigma^\+ Q \d_\beta\Sigma)=(1/3)\tr [t^3 (L_\beta+R_\beta)]$.  
The WZW action is zero in the
absence of the external fields: $S_\textrm{WZW}[0]=0$.  
In the presence of external fields, it
becomes
\begin{equation}
  S_{\rm WZW} = \int\!d^4x\, \left\{
  -A^B_\mu j_B^\mu + 
%  \frac{\epsilon^{\mu\nu\alpha\beta}}{48\pi^2}
  \frac{\epsilon^{\mu\nu\alpha\beta}}{16\pi^2}
\left(\frac13
eA_\mu
  \tr(L_\nu L_\alpha L_\beta) 
%   - \frac{ie^2}{32\pi^2}\epsilon^{\mu\nu\alpha\beta}
-  \frac{ie^2}{2}
F_{\mu\nu}A_\alpha
  \tr [t^3 (L_\beta+R_\beta)]
\right)
\right\},
\end{equation}
and
\begin{align}\label{jB-2fl}
  j^\mu_B = -\frac1{24\pi^2}\epsilon^{\mu\nu\alpha\beta}
\left\{ 
     \tr(L_\nu L_\alpha L_\beta) 
%+ \frac{ie}{8\pi^2}
- {3ie}
\d_\nu \left[
%     \epsilon^{\mu\nu\alpha\beta} 
A_\alpha
     \tr(t^3 L_\beta + t^3 R_\beta)\right]
\rule{0pt}{1em}\right\},
\end{align}
or
\begin{equation}
  \label{eq:jB-gw2}
   j^\mu_B = - \frac1{24\pi^2}\epsilon^{\mu\nu\alpha\beta}
\left\{
    \tr[(\Sigma D_\nu\Sigma^\+) (\Sigma D_\alpha\Sigma^\+) 
    (\Sigma D_\beta\Sigma^+)]
%  + \frac{ie}{16\pi^2}\epsilon^{\mu\nu\alpha\beta}
-\frac{3ie}{2}
F_{\nu\alpha}
  \tr[t^3(\Sigma D_\beta\Sigma^+ + D_\beta\Sigma^\+\Sigma)]
\right\}.
\end{equation}
The WZW action can therefore be written as
\begin{equation}\label{eq:Q=B/2}
  S_{\rm WZW} = -\int\!d^4x\, \left( A^B_\mu + \frac e2 A_\mu\right) j^\mu_B.
\end{equation}
The second term is the contribution of the baryon charge to the
electric charge of a baryon as in the Gell-Mann-Nishijima formula
 $Q=I_3+N_B/2$.

Consider one particular case, when  $\Sigma$ is restricted to the form
\begin{equation}
  \Sigma = \exp\left(\frac i{f_\pi} \tau_3 \varphi_3 \right),
\end{equation}
and the external field is chosen to be a constant magnetic field
$B_i=\epsilon_{ijk}F_{jk}/2$ and baryon chemical potential
$A_\nu^B=(\mu,\bm{0})$.  In this case the WZW action assumes an even simpler
form [only the last term in Eq.~(\ref{jB-2fl}) survives]:
\begin{equation}\label{eq:Sphi3}
  S_{\rm WZW} = \frac e{4\pi^2f_\pi}\int\! d^4x\,
  \mu \bm{B} \cdot \bm{\nabla}\varphi_3.
\end{equation}
This form of the magnetic effective action has been written down and
discussed in Ref.~\cite{Son:2004tq}, where it was interpreted
as a nonzero magnetization of a $\pi^0$ domain wall at finite
$\mu$ given by
\begin{equation}
  \label{eq:magnetization}
 \bm M = \frac e{4\pi^2f_\pi} \mu  \bm{\nabla}\varphi_3.
\end{equation}
In this paper  we point out that the same term is responsible for the
nonzero baryon density of a domain wall in an external magnetic field:
\begin{equation}
  \label{eq:baryondensity}
  n_B= \frac e{4\pi^2f_\pi}
  \bm{B} \cdot \bm{\nabla}\varphi_3.
\end{equation}

\section{$\pi^0$ domain wall in a magnetic field}
\label{sec:pi0-domain-wall}

\subsection{Local stability}
\label{sec:local-stability}

To treat the $\pi^0$ domain wall and the fluctuations around it, it is
most convenient to use the following parametrization
\begin{align}\label{param}
  \sigma = f_\pi \cos\chi \cos\theta,\qquad & \pi^1 = f_\pi\sin\chi\cos\phi,\\
  \pi^0 = f_\pi \cos\chi \sin\theta,\qquad & \pi^2 = f_\pi\sin\chi \sin\phi.
\end{align}
The Lagrangian (without the magnetic field) is given by
\begin{equation}\label{eq:Lpi0}
  {\cal L} = \frac{f_\pi^2}2[ (\d_\mu\chi)^2 + \cos^2\chi (\d_\mu\theta)^2
  + \sin^2\chi(\d_\mu\phi)^2 ] - f_\pi^2 m_\pi^2(1-\cos\chi\cos\theta).
\end{equation}
The $\pi^0$ domain wall corresponds to the following static solution
to the field equations,
\begin{equation}\label{kink}
  \chi =0, \qquad \theta = 4\arctan e^{m_\pi z}.
\end{equation}

Topologically, since Eq.~(\ref{kink}) corresponds to a contractible loop
in the SU(2) group manifold (S$^3$),
the wall can be ``unwound.'' Moreover, in the absence of a magnetic field 
the $\pi^0$ domain wall is not even {\em locally} stable.
This can be seen by analyzing small fluctuations around the
solution (\ref{kink}).  For small $\pi_1$ and $\pi_2$ the Lagrangian
is given by
\begin{equation}
  {\cal L} = \frac12[(\d_\mu\pi_1)^2 +(\d_\mu\pi_2)^2] -
    \frac{m_\pi^2}2\left(
    1 - \frac6{\cosh^2 m_\pi z}\right)(\pi_1^2+\pi_2^2).
\end{equation}
The equations of motion are
\begin{equation}
  -(\d_x^2 + \d_y^2)\pi^a - \d_z^2 \pi^a + m_\pi^2 \left( 
     1 - \frac6{\cosh^2 m_\pi z}\right) \pi^a = E^2\pi^a.
\end{equation}
The corresponding Schr\"odinger equation has two bound states.  The
lowest state is tachyonic,
\begin{equation}
  E^2 = k_x^2 + k_y^2 - 3 m_\pi^2\,,
\end{equation}
so the wall is locally unstable. (The second bound state corresponds to 
a zero mode of the wall.)

In the magnetic field, the Laplacian in the $(x,y)$ plane becomes the
Hamiltonian of a particle in a magnetic field, whose spectrum
(the Landau levels) is well known, leading to
\begin{equation}\label{eq:E2=eB-3m2}
  E^2 = (2n+1) eB - 3 m_\pi^2,\quad n=0,1,\ldots
\end{equation}
Therefore, when the magnetic field exceeds the value
\begin{equation}
  B_0 = \frac{3 m_\pi^2}e \approx 1.0\times 10^{19}~\textrm{G},
\end{equation}
the $\pi^0$ domain wall becomes locally stable.

\subsection{Global stability at finite $\mu$}
\label{sec:stab-fin-mu}

Substituting the configuration (\ref{kink}) into the Lagrangian
(\ref{eq:Lpi0}), one finds 
the following energy density per unit area,
\begin{equation}
  \label{eq:Energy/S}
  \frac{\cal E}{S} = 8f_\pi^2 m_\pi.
\end{equation}

At finite baryon chemical potential $\mu$ and  in the presence of 
a magnetic field $F_{xy}=B$ (i.e., $B_z=-B$), the configuration 
(\ref{kink}) carries a baryon number
according to Eq.~(\ref{eq:baryondensity}) with $\varphi_3=f_\pi\theta$.
The baryon number per unit surface area is thus given by
\begin{equation}\label{rhoB}
  \frac{N_B}{S} = \frac{eB}{2\pi}\,.
\end{equation}
Being a total derivative, the WZW term (\ref{eq:Sphi3}) does not
affect the field equations.

The energy per baryon number of the $\pi^0$ domain wall is
\begin{equation}\label{eq:E/N-wall}
  \frac{\cal E}{N_B} = 16\pi\frac{f_\pi^2 m_\pi}{eB}\,.
\end{equation}
When the baryon chemical potential exceeds the value of that ratio,
i.e., for  $\mu>16\pi{f_\pi^2 m_\pi}/{(eB)}$,
the wall becomes energetically more favorable than the vacuum,
and the ground state must be a stack of parallel domain walls,
(at least) as long as $\mu\lesssim m_N$---the energy per baryon number
of the nuclear matter. In order to be more favorable than
the nuclear matter at $\mu\approx m_N$ the ratio (\ref{eq:E/N-wall})
must be less than $m_N$. This happens if the magnetic field exceeds
\begin{equation}
  B_1 = \frac{16\pi f_\pi^2 m_\pi}{e m_N} \approx 1.1 \times
  10^{19}~\textrm{G}.
\end{equation}
In the chiral limit $m_\pi\to0$,  $B_1\gg B_0$, 
but for the real-world pion mass
$B_1$ is only slightly higher than $B_0$.

According to Eq.~(\ref{eq:Q=B/2}),
the $\pi^0$ domain wall carries a finite surface electric charge density
equal to a half of the baryon charge density given by Eq.~(\ref{rhoB}).  
Within QCD, this charge
can be neutralized by the $\pi^-$ bosons localized on the wall: 
according to Eq.~(\ref{eq:E2=eB-3m2}) the energy cost of adding
a~$\pi^-$ vanishes at $B=B_0$. The number of charged pions
necessary to neutralize the wall 
fills exactly a half of the first Landau level.  
This suggests that the electrically neutral ground state may show
quantum Hall behavior.
For $B>B_0$, each pion cost an energy of $(e(B-B_0))^{1/2}$.
However, for $B>B_0$, within the full Standard
Model (with electromagnetism), other mechanisms of neutralizing the
electric charge of the wall may compete with adding charged pions 
(e.g., adding electrons).  Since the energy of adding one electron
to the system is only $m_e$ (its lowest Landau level energy), our
estimate for $B_1$ is largely unaffected.

\subsection{Structure and baryon charge of a finite domain wall}
\label{sec:struct-finite-wall}

So far we have considered an infinite domain wall.  Let us now
consider a large, but finite-size, domain wall. For the infinite
wall, the baryon charge, given by Eq.~(\ref{rhoB}), comes from the {\em
  second term} in the baryon current (\ref{jB-2fl}), which gives
Eq.~(\ref{eq:Sphi3}). This term is a full derivative, so for a
{\em finite} wall it must vanish. Where does the baryon number
come from in this case? We now demonstrate explicitly
that the finite domain wall carries a baryon number that comes from 
the first term in Eq.~(\ref{jB-2fl}).

We consider a flat domain wall with a circular boundary.  We use
cylindrical coordinates $(\rho,\varphi,z)$ with the origin at the
center of the wall.  The boundary of the wall is chosen to be
$z=0$, $\rho=R$.  We assume the radius $R$ is much larger than the
thickness of the wall, $R\gg m_\pi^{-1}$.

We use the parametrization~(\ref{param}).  We expect that when
$\rho<R$ and $R-\rho>m_\pi^{-1}$, we are sufficiently far away from
the boundary so that the domain wall is given by Eq.~(\ref{kink}).  In
particular, when $z$ varies from $-\infty$ to $+\infty$, $\theta$
jumps by $2\pi$:
\begin{equation}
  \theta(z=+\infty) - \theta(z=-\infty) = 2\pi, \qquad \rho<R.
\end{equation}
When $\rho>R$, one does not cross any domain wall as one moves along
the $z$ direction,
\begin{equation}
  \theta(z=+\infty) - \theta(z=-\infty) = 0, \qquad \rho>R.
\end{equation}
We find that $\theta$ is a multiple-valued function: it changes
by $2\pi$ when we move along a small loop around the boundary
$\rho=R$, $z=0$.  To avoid a singularity in the fields themselves,
$\cos\chi$ has to vanish on the boundary.  We can choose
\begin{equation}
  \chi(\rho=R,z=0) = \frac\pi 2\,.
\end{equation}

We expect that $\chi$ is nonzero only near the boundary.  So the
$\pi^1$ and $\pi^2$ fields differ substantially from 0 only near
$\rho=R$.  As these fields describe the charged pions, the boundary of the
domain wall is a superconducting string~\cite{Witten:1984eb}.  At the
boundary $\rho=R$, the charged pion condensate is largest,
$(\pi^1)^2+(\pi^2)^2=f_\pi^2$.  Moreover, the phase $\phi$ of the
charged pion condensate has a nontrivial winding number around the
circle $\rho=R$.  Indeed, in order to minimize the kinetic energy, this
winding number is equal to the magnetic flux that goes through the
contour, in unit of the elementary flux:
\begin{equation}
  \phi(\varphi=2\pi) - \phi(\varphi=0) = \frac1{2\pi}eB(\pi R^2)
  = \frac12 eB R^2 .
\end{equation}
Because of continuity, the phase $\phi$ has the same winding number on any
contour that surrounds the $z$ axis, $\rho=0$.  To avoid singularity on
this axis, we must have $\sin\chi=0$ at $\rho=0$.  We choose
$\chi(\rho=0)=0$.

Thus we find that a finite $\pi^0$ domain wall has a peculiar feature:
the phase $\phi$ makes $\frac12 eBR^2$ full circles on any contour
that surrounds the axis $z=0$, and the phase $\theta$ makes a full
circle on any contour that has linking number one with the boundary
$\rho=R$ of the wall.  The phase $\chi$ changes from $0$ on the $z$
axis to $\pi/2$ on the boundary of the wall.  It is easy to see that
the configuration has the topology of a Skyrmion with the baryon
charge $N_B=\frac12 eBR^2$.  In can be already seen from Eq.~(\ref{jB-gi})
but it is instructive to check that Eq.~(\ref{jB-cons}) gives the same
result.  Indeed, the full derivative term in Eq.~(\ref{jB-cons}) does
not contribute to the total baryon charge and we have
\begin{equation}
  N_B = -\frac1{24\pi^2}\int\!d^3x\, \epsilon^{ijk} \tr(L_i L_j L_k).
\end{equation}
Changing coordinate system to $\chi$, $\theta$, and
$\phi$, one finds that the baryon charge is equal to $\frac12 eB R^2$.
The baryon charge per unit surface area is the same as in
Eq.~(\ref{rhoB}).

\section{Color superconducting phases}
\label{sec:colorsup}

So far, we have considered the effect of the magnetic field on
low-density matter.  In this Section, we consider the effect of the
magnetic field on the structure of high-density quark matter.  Such
high-density matter may exist in one of the color-superconducting
phases (see, e.g., 
Refs.~\cite{Rajagopal:2000wf,Schafer:2003vz,Alford:2006wn,Shovkovy:2007zz,Alford:2007xm} for
reviews).  We shall see that due to the existence of light pseudoscalar
Nambu-Goldstone bosons, stacks of domain walls for
such bosons can be generated, and because the corresponding bosons are
light, the critical magnetic field can be much lower than in 
vacuum.

\subsection{2SC phase in a magnetic field}
\label{sec:2sc-phase-magnetic}

Theoretically, the simplest color superconducting phase is the
two-flavor superconducting (2SC) phase~\cite{Alford:1997zt,Rapp:1997zu}. 
On the phase diagram, this phase
occupies a window of chemical potential next to 
low-density nuclear matter: right after the chiral symmetry is
restored, but before the density of strange quarks becomes
significant.

In this regime, the attraction between quarks in the color-triplet
mutual state leads to an instability of the Fermi surface
due to the familiar Cooper mechanism.  The resulting Cooper pair condensate
has the quantum numbers of a color triplet and an isospin singlet, and carries
zero angular momentum.

Perturbatively, there are two such condensates: the left- and the 
right-handed quark pairs: $X\sim q_Lq_L$ and $Y\sim q_Rq_R$. The 
gauge-invariant (color singlet) order parameter is the singlet made out
of $X$ and $Y$ color vectors: $\Sigma=XY^\dagger$. Like $X$ and $Y$,
$\Sigma$ is also an isosinglet: the isospin SU(2)$_L\times$
SU(2)$_R$ chiral symmetry
is not broken in the 2SC phase. However, since the phases of $X$ and $Y$ change
in opposite directions under the axial isospin singlet U(1)$_A$ symmetry, 
the phase of the
order parameter $\Sigma=XY^\dagger$ changes under U(1)$_A$. This means
that the U(1)$_A$ symmetry is broken by the condensate.

In reality, this U(1)$_A$  symmetry is not a true symmetry of
QCD---it is violated by the quantum fluctuations of the
gluon fields via an anomaly. However, the vacuum configurations of the
gluon fields responsible for this violation, i.e., the instantons,
are suppressed at large baryon density due to color Debye screening,
and the U(1)$_A$ transformation can be treated as an approximate symmetry 
at large $\mu$.

In the 2SC phase, where the U(1)$_A$ is spontaneously broken, 
the measure of the explicit violation of this symmetry by
anomaly/instantons is the mass $m_\eta$ of the Goldstone boson 
(which we call $\eta$).
This mass decreases very fast with $\mu$ 
(see below and Ref.~\cite{SSZ}). The
smallness of $m_\eta$ is what is responsible for the low value of the
critical magnetic field.

The effective Lagrangian density for the $\eta$ boson in the 2SC phase is
\cite{SSZ}
\begin{equation}
  {\cal L} = f^2 [(\d_0\varphi)^2 - u^2 (\d_i\varphi)^2 -
       m_\eta^2(1-\cos\varphi)]\, ,
  \label{LeffV}
\end{equation}
%\begin{equation}
%  L = f^2 [(\d_0\varphi)^2 - u^2 (\d_i\varphi)^2] -
%     V_{\rm inst}(\varphi) \, .
%  \label{LeffV}
%\end{equation}
where $\varphi$ is the local value of the U(1)$_A$ phase whose
fluctuations generate Goldstone boson $\eta$. For asymptotically 
large $\mu\gg\Lambda_\textrm{QCD}$ the low-energy constants 
in the effective Lagrangian (\ref{LeffV}) are
calculable~\cite{Beane:2000ms,inverse-ordering}:
\begin{equation}\label{eq:f,u}
  f^2 = \frac{\mu_q^2}{8\pi^2} \, , \qquad  u^2 = \frac13 \, .
\end{equation}
and
% \begin{equation}
%   V_{\rm inst}(\varphi) = a \mu^2\Delta^2(1-\cos\varphi) \, ,
%   \label{Vinst}
% \end{equation}
\begin{equation}
   m_\eta = \sqrt{\frac a2} \, \frac{\mu_q}{f} \Delta 
   = 2\pi \sqrt a \Delta \, ,
   \label{meta}
\end{equation}
where $\Delta$ is the superconducting gap and $a$ has been 
estimated in Ref.~\cite{SSZ}
\begin{equation}
  a = 5 \times 10^4 \biggl(\ln\frac{\mu_q}{\LambdaQCD}\biggr)^7
  \biggl(\frac{\LambdaQCD}{\mu_q}\biggr)^{29/3} \, .
  \label{amu}
\end{equation}
In Eqs.~(\ref{eq:f,u})---%, (\ref{meta}), and 
(\ref{amu}), $\mu_q$
denotes the quark chemical potential: $\mu_q\equiv\mu/3$.

The domain wall configuration $\varphi=4\arctan[\exp(m_\eta z/u)]$
is a static solution of the equations of motion with energy
per unit surface area given by
\begin{equation}
  \frac {\cal E}S = 16 u f^2 m_\eta
%8\sqrt{2a}\, uf\mu\Delta  
% {4u\over\pi}\sqrt{a}\mu^2\Delta 
  \, .
  \label{sigma}
\end{equation}
Unlike the $\pi^0$ domain wall in Section~\ref{sec:pi0-domain-wall},
it is locally stable because of the topology of U(1)$_A$: 
the wall can be unwound only by changing the
magnitude of $\Sigma$, which requires energies beyond the scale of
the effective Lagrangian (\ref{LeffV}).

The interaction of $\varphi$  with the magnetic field due to the axial anomaly
is described by \cite{Son:2004tq}
\begin{equation}
  \label{eq:L-B-phi}
{\cal L} = %\frac{e\mu}{24\pi^2}
\frac{e\mu}{36\pi^2}
\,\bm {\nabla}\varphi \bm {\cdot B}.
\end{equation}
Being a total derivative, this term does not change the 
field equations for
$\varphi$, but it does contribute to the total 
free energy of a domain wall. In
particular, for the domain wall
perpendicular to the
homogeneous field $\bm B$, the magnetic free energy per unit area 
is given by $e\mu
B/(18\pi)$, which can be interpreted as the surface density of dipole
magnetic moment directed perpendicularly to the wall,
\begin{equation}
  \label{eq:mag-moment}
  \frac{|\textswab m|}{S}=\frac {e\mu}{18\pi} \, . %{12\pi}.
\end{equation}

For sufficiently large $B$, the free energy gain due to the interaction of
the wall with the magnetic field outweighs the surface energy cost of
creating a wall (\ref{sigma}).
Thus the critical field is
\begin{equation}
  \label{eq:B-crit}
  B_c=\frac{\cal E}{|\textswab{m}|}=288 %192
 \pi u \frac{f^2m_\eta}{e\mu} 
= 
\frac{4}{\sqrt3\pi}\frac{\mu m_\eta}{ e} 
\approx 1.2\cdot 10^{18}\textrm{ G} \times
\left(\frac\mu{1 \textrm{ GeV}}\right)
\left(\frac {m_\eta}{10 \textrm{ MeV}}\right).
\end{equation}

For $B>B_c$, the domain walls are energetically favorable and
(provided boundary conditions allow) they will stack up until their
mean separation is of the order of their width $1/m_\eta$.

For comparison, the critical magnetic field needed to destroy 
superconductivity is at least of order $\mu\Delta/e$~\cite{Alford:1999pb}.
Due to fast descrease of $m_\eta$ with $\mu$, the value of $B_c$ is
much lower than the critical field at large $\mu$.

%The magnetization (magnetic moment per unit volume) 
%at this point is of order $e\mu m/(6\pi)$, and continues to grow
%roughly linearly with $B$.
%It then continues to grow
%linearly with $B$: $M=\chi B$, with $\chi=e^2/(12\pi^2)$.

\subsection{CFL}
\label{sec:cfl}

At large $\mu$ one eventually enters the regime where
the mass of the strange quark can be neglected, the density of
strange quarks is as large as that of up and down and the pairing
involving all three flavors becomes energetically favorable.  This
pairing state is called color-flavor-locked
(CFL) phase~\cite{CFL}.

In the CFL phase, the Cooper pairs are both flavor and color
triplets, i.e., $X\sim q_Lq_L$ and $Y\sim q_Rq_R$ each
carry a color and a flavor index and transform as color-flavor matrices 
$X\to LXC^T$ and $Y\to RYC^T$ under the flavor and color 
SU(3)$_L\times$SU(3)$_R\times$SU(3)$_C$ transformations.
The gauge-invariant order parameter $\Sigma=XY^\dagger$
transforms in the same way as the ordinary chiral condensate in vacuum,
$\Sigma\to L\Sigma R^\dagger$. Therefore the chiral
SU(3)$_L\times$SU(3)$_R$ is broken, in the CFL phase, down to the
vector-like SU(3)$_{L+R}$ as it is in the vacuum.

Similarly to
the 2SC phase, the U(1)$_A$ symmetry is also spontaneously broken in
the CFL phase. The SU(3)$_L\times$SU(3)$_R\times$U(1)$_A$ symmetry is 
explicitly violated by instantons and quark
masses, so all Nambu-Goldstone bosons are massive.  For simplicity,
we consider the regime reached at asymptotically high $\mu$ where one
can neglect the contribution of instantons to all masses.
%Also similarly to 2SC, the anomalous violation of this
%symmetry by instantons is vanishingly small at large $\mu$.
%Turning on quark masses lifts up the masses of all the Goldstone
%bosons of the SU(3)$_L\times$SU(3)$_R\times$U(1)$_A$ symmetry. 
The lightest Nambu-Goldstone 
boson in this case  is an isosinglet
which has the quantum number of $\bar s s$, i.e., a mixture of $\eta$ 
and $\eta'$~\cite{inverse-ordering}.  Its mass square is 
given
by~\cite{inverse-ordering}
\begin{equation}
  m_{\bar ss}^2 = \frac{3\Delta^2 m_u m_d}{\pi^2f^2}
\end{equation}
where $f^2\sim\mu^2$ is given below in Eqs.~(\ref{ff1f8}) and (\ref{f1f8}).

The effective Lagrangian for this field, $\varphi_{\bar ss}$,
is similar to the Lagrangian (\ref{LeffV}), 
\begin{equation}
  {\cal L} = f^2 [(\d\varphi_{\bar ss})^2 - u^2 (\d_i\varphi_{\bar ss})^2
   -m_{\bar ss}^2 (1-\cos\varphi_{\bar ss})].
\end{equation}
Since the boson is a mixture of the $\eta$ and $\eta'$, its decay constant 
is a linear combination of the singlet and the octet decay constants.  
One can easily derive
\begin{equation}\label{ff1f8}
  f^2 = \frac1{12} (f_{\eta'}^2 + 2 f_\pi^2),
\end{equation}
where $f_\eta'^2$ and $f_\pi^2$ have been computed in
Ref.~\cite{inverse-ordering},
\begin{equation}\label{f1f8}
  f{_\eta'}^2 = \frac 34 \frac{\mu_q^2}{2\pi^2}, \qquad
  f_\pi^2 = \frac{21-8\ln 2}{18} \frac{\mu_q^2}{2\pi^2}\,.
\end{equation}

%for $\eta$ with
%replacements $m_\eta\to m_\eta'$ and $\varphi\to\varphi'$. The mass of
%$\eta'$ is given by a formula similar to Eq.~(\ref{meta})and (\ref{amu})
%with the coefficient $a'$, replacing $a$, smaller by a factor of order
%$m_s/\mu_q$~\cite{SSZ}. 

The anomalous coupling of the $\varphi_{\bar ss}$ field to the
magnetic field and baryon chemical potential is given
by~\cite{Son:2004tq}
\begin{equation}
  \label{eq:L-B-phi-prime}
{\cal L}' = %\frac{e\mu}{24\pi^2}
\frac{e\mu}{12\pi^2}
\,\bm {\nabla}\varphi_{\bar ss} \bm {\cdot B}.
\end{equation}

Therefore the critical magnetic field in CFL can be estimated
as
\begin{equation}
  \label{eq:H_c-cfl}
  B_c' = 96\pi u \frac{f^2 m_{\bar ss}}{e\mu}
       = \frac{111-32\ln2}{81\sqrt 3\pi} \frac{\mu m_{\bar ss}}e
       = \frac{8\sqrt{111-32\ln2}}{3\sqrt6\pi} \Delta\sqrt{m_u m_d}\,.
%       = 1.5 \cdot 10^{17}\textrm{ G} \times
%\left(\frac\mu{1.5 \textrm{ GeV}}\right)
%\left(\frac {m_{\bar ss}}{3 \textrm{ MeV}}\right).
\end{equation}
Numerically, it can be written as
\begin{equation}
  B_c' = 1.0 \cdot 10^{17}\textrm{ G} \times
  \left(\frac\mu{1.5 \textrm{ GeV}}\right)
  \left(\frac {m_{\bar ss}}{2 \textrm{ MeV}}\right)
  = 8.3\cdot 10^{16}\textrm{ G} \times 
    \left( \frac\Delta {30 \textrm{ MeV}}\right)
    \left(\frac{\sqrt{m_u m_d}}{5 \textrm { MeV}}\right).
\end{equation}

Numerically, the value obtained here is close to the theoretical 
upper limit of magnetic fields possible in neutron stars~\cite{Duncan:1992hi}.

\section{Ferromagnetic quark matter}
\label{sec:ferro}

The presence of the anomaly term $\mu \bm\nabla\varphi\bm{\cdot B}$ in the
Lagrangian implies that if a gradient of a pseudoscalar boson is
spontaneously generated in the ground state, then the state will carry a
spontaneous magnetization  proportional to
$\mu\bm\nabla\varphi$---i.e., it will be ferromagnetic.\footnote{The
ferromagnetism of an axial domain wall in vacuum has been discussed in
Refs.~\cite{Iwazaki:1996xf,Cea:1998ep} using a microscopic
approach in connection with the primordial magnetic field generation
(see also Ref.~\cite{Forbes:2000gr}). It is
worth pointing out that unlike the vacuum case, where the magnetization is
forbidden by $C$ parity~\cite{Voloshin:2001iq}, in the case we consider 
the $C$ parity is
explicitly broken by the background baryon charge density.}
Such a phase has been discussed in the literature
under the name ``Goldstone boson current'' or ``supercurrent''
phase. This phase becomes
favorable in the range of chemical potentials between CFL and 2SC
phases. If we start from the CFL phase and decrease the chemical potential
$\mu$, the splitting of the Fermi surfaces, $m_s^2/(2p_F)$, 
caused by strange quark mass
$m_s$ leads to an instability~\cite{Casalbuoni:2004tb}.
A similar instability occurs in the 2SC phase~\cite{Huang:2004bg}.
% of the CFL ground state toward
%condensation of Cooper pairs with {\em non-zero} wave-vector. From the point
%of view of effective theory this is the Goldstone boson current state.

In the language of the effective theory (chiral perturbation theory
with baryon excitations~\cite{Kryjevski:2004jw}), the instability
arises when a fermion excitation mode (a baryon) is about to turn
gapless~\cite{Alford:1999xc,Alford:2003fq} 
due to the effective chemical potential, $m_s^2/(2p_F)$,
introduced by the strange quark mass. Because of the existence of a
bilinear coupling $\bm\nabla\varphi\bm\cdot\bm j$ of the
``supercurrent'' $\bm\nabla\varphi$ of a Goldstone boson to the normal
current $\bm j=\psi^\dag\bm v\psi$ of the fermion $\psi$, when the
fermion is nearly gapless one can lower the energy by simultaneously
generating the Goldston boson current $\bm\nabla\varphi$ and the
ordinary current $\bm j$ of opposite
directions~\cite{Son:2005qx,Kryjevski:2005qq,Schafer:2005ym}.

For definiteness,
we shall discuss the Goldstone boson current state in the kaon-condensed CFL
phase (CFLK$^0$)~\cite{Gerhold:2006np}.  Most of the discussion is 
also relevant
for the current phase in the CFL phase without kaon
condensation~\cite{Gerhold:2006dt} and in the 2SC phase~\cite{Huang:2005pv}.

As discussed in Ref.~\cite{Gerhold:2006dt}, to leading order in
the strong-coupling constant $\alpha_s$, there is a degeneracy between
the ``vector current'' state and the ``axial current'' state.  In the
vector current state $X$ and $Y$ rotate in the same direction as one
moves along the $z$ direction, and in the axial current state they
rotate in the opposite directions.  We shall assume that the axial current
state is favored.  In this state, the gauge invariant order parameter
$\Sigma$ varies in space.

We should stress that the term ``current state'' is somewhat
misleading, as the total current in the ground state is zero.  For
example, in the axial current state the axial current from the
condensate is compensated by the axial current of gapless fermions.
However, in contrast to the conserved currents, there is no reason for
the {\em magnetization} to vanish.

According to Ref.~\cite{Gerhold:2006np}, the Goldstone boson
current CFLK$^0$ phase appears when the effective chemical potential
 $\mu_s$ induced by the strange quark mass is in a
narrow range
\begin{equation}
  \label{eq:mus-range}
   1.605\Delta <  \mu_s\equiv \frac{m_s^2}{2p_F} <1.615\Delta.
\end{equation}
 Here $p_F=\mu/3$ is the quark Fermi momentum.

The chiral field $\Sigma$ in the  CFLK$^0$ phase is
\begin{equation}
\Sigma = \exp(-i cz Q) \exp\left( \frac{i\pi}2 \lambda_6\right)\exp(-i cz Q)
= \exp(-i 2 cz Q)\exp\left(\frac{i\pi}2 \lambda_6\right),
\end{equation}
where $c$ is some constant that is determined by energy minimization.
There is also a U(1)$_A$ linear background but it does not contribute
to the anomaly that we need (since $\tr Q=0$).  It turns
out~\cite{Gerhold:2006np} that the minimum of the energy is achieved when
$c\approx\Delta$, so one is stretching the applicability of the
effective theory.  We are interested in rough estimates, so we shall
use the effective theory extrapolation.  In the ground state,
\begin{equation}
  \Sigma\d_z \Sigma^\+ = \d_z \Sigma^+ \Sigma = 2icQ,
\end{equation}
so the WZW term contribution to the Lagrangian is
\begin{equation}
  \frac e{2\pi^2} \mu B \tr(c Q^2) = \frac e{3\pi^2} \mu B c.
\end{equation}
Putting $c=\Delta$, we find the magnetic moment density (magnetization)
\begin{equation}
\label{eq:M}
  M = \frac e{3\pi^2} \mu\Delta = 2.4 \cdot 10^{16}~\textrm{G}\times
     \left(\frac\mu{1.5 \textrm{ GeV}}\right)
     \left(\frac\Delta{30\textrm{ MeV}}
     \right).
\end{equation}

An important point not to be overlooked in such a calculation of the
magnetization is a possible contribution of the near-gapless fermions that are
present in the system. In the particular case of CFLK$^0$ considered
here, these fermions are electrically neutral and do not contribute.

%The typical size of the magnetic field generated is therefore of order
%$10^{16}~\textrm{G}$.  It is interesting that the value of the field
%is roughly of the same order of magnitude as the fields in magnetars.

What is a typical value of the magnetic field generated by this
mechanism inside a neutron or quark star? The local baryon chemical
potential is a function
of the distance to the center of the star and is 
increasing towards the center of the star. Let us
assume that it reaches the narrow range in which the Goldstone boson
current CFLK$^0$ phase appears~\cite{Gerhold:2006np}
\begin{equation}
  \label{eq:mu-range}
 \frac{m_s^2}{2\Delta} (1.615)^{-1} <\frac\mu3<  
 \frac{m_s^2}{2\Delta} (1.605)^{-1},
\end{equation}
before reaching the maximum at the star's center.  This range maps onto
a relatively thin shell inside the star, and we denote its mean radius
as $R$ and the thickness $d$ (we estimate below $d\sim 100~\textrm{m}$ for a
typical star of $R_*\sim 10$ km radius). Assuming that the magnetization in the
shell is uniform, one finds that the magnetic field it creates
outside is the same as that of a dipole moment equal to the total
magnetic moment of the shell $M\cdot 4\pi R^2 d$. 
Near the surface of the shell this field is of order
\begin{equation}
  \label{eq:H-vs-M}
  B  \sim  M \,\frac{d}{R}
\end{equation}
(within the shell the field is much larger $B\sim M$ and it is
zero inside the non-ferromagnetic region surrounded by the shell
-- the shell screens the field out of it).
% and dipole-like outside
%that shell we can estimate the energy of this magnetic field as
%$H^2\cdot R^3$, while the contribution of the ferromagnetic shell as
%$-MH\cdot R^2d$. The total energy is minimized when
From Eq.~(\ref{eq:mu-range}) the width of the range in $\mu$
is of the order of 10 MeV. Taking the typical range of
variation of $\mu$ in the star of order 500 MeV, we estimate 
 $d/R\sim 10/500=0.02$. Using
the estimate (\ref{eq:M}) for the magnetization $M$, 
we find from (\ref{eq:H-vs-M})
that typical fields generated by such mechanism are of order
$B\sim 10^{14}-10^{15}$ G, which is the right order of magnitude to account for
the observed magnetic fields of magnetars.  
%However, the mechanism
%considered here has problem explaining fields of order $10^{16}~\textrm{G}$
%if they exist:
%such fields seem to require the current phase to exist in a window
%larger than in Eq.~(\ref{eq:mus-range}).

\section{Conclusion}

In this paper we discussed the effects of the magnetic field
on the ground state of QCD at different values of baryon density.
The key mechanism which leads to the effects we describe is due
to the axial anomaly. In the effective low-energy description
of QCD -- the chiral Lagrangian for the Goldstone bosons --
this effect is represented by a term which appears when we
gauge the topological (Goldstone-Wilczek) baryon
current. On the microscopic level, it is given by the triangle diagram
with the baryon, electromagnetic and axial charge currents at the
vertices.

We have demonstrated that in a sufficiently strong magnetic field the
most stable state with finite baryon number is not nuclear matter, but
a $\pi^0$ domain wall. Similarly, at higher baryon densities, the most
stable state in a sufficiently strong magnetic field is that of an
isoscalar axial ($\eta$ or $\eta'$) domain wall.

We also show that the states of quark matter with Goldstone boson
current are ferromagnetic, and show that their magnetization is
related to triangle anomalies.  We estimate the magnetic field
generated by such a mechanism in a typical neutron/quark star to be of order
$10^{14}-10^{15}~\textrm{G}$, which is a relevant magnitude for
neutron star phenomenology.

Further work is needed to understand if such ferromagnetic quark matter
exists.  In particular, one should understand whether the ``vector
current'' or ``axial current'' state is favored.  In addition, one
should determine if the current states are favored compared to other
candidate ground states (for example, the Fulde-Ferrell-Larkin-Ovchinnikov 
states with multiple plane waves)~\cite{Alford:2007xm}.

\acknowledgments

We thank T.~D.~Cohen, D.~B.~Kaplan, S.~Reddy and M.~Voloshin for discussions.
D.T.S. is supported, in part, by DOE grant No.\ DE-FG02-00ER41132.
M.A.S. is supported, in part, by DOE grant No.\ DE-FG02-01ER41195.

\end{document}